# A Lockable ERC20 Token for Peer to Peer Energy Trading

Liana Toderean, Claudia Antal, Marcel Antal, Dan Mitrea, Tudor Cioara, Ionut Anghel, Ioan Salomie

Computer Science Department
Technical University of Cluj-Napoca
Cluj-Napoca, Romania
{liana.toderean, claudia.pop, marcel.antal, dan.mitrea, tudor.cioara, ionut.anghel, ioan.salomie}@cs.utcluj.ro

*Abstract* — **In this paper, we address the digitization of physical assets using blockchain technology focusing on energy and peer-to-peer trading on decentralized energy markets. Because they are forward markets and operate on a day-ahead timeline, the energy transactions are settled only at the movement of energy delivery. Having the option of locking the energy tokens by a third-party escrow becomes a highly desirable feature. Thus, we define a Lockable ERC20 token that provides the option for an owner to lock some of its tokens using smart contracts. A time-lock and an escrow party account or smart contract can be specified allowing the tokens to be unlocked when certain business conditions are met. For validation purposes, we have considered a peer-to-peer energy trading scenario in which the Lockable ERC20 token was used to digitize the surplus of energy of prosumers. In this case, the energy tokens committed in blockchain transactions are successfully locked up until the actual delivery of energy, the settlement considering the monitored data of energy meters.**

## I. INTRODUCTION

Blockchain is a technology that is gaining a lot of attention, bringing a lot of potential benefits in smart grid management scenarios such as the tamper-proof registration of transactions and peer-to-peer trading of energy as a digital asset. Other advantages are in terms of resiliency, adaptability, fault tolerance, security, and trust. In fact, peer-to-peer trading enables market participants (i.e., prosumers) to trade their excess energy or reducing their demand leveraging on flexibility [1].

Trading energy in a peer-to-peer fashion poses significant challenges from a technological, energy management, and regulation perspective. From a technological point of view, a fundamental aspect is the representation of energy as a digital asset using tokens to enable its trading. The generation of energy tokens is usually done proportional to the forecasted energy excess or flexibility of a prosumer, but this process is exposed to uncertainty depending on parameters such as local weather conditions. This can lead to sub-optimal situations in which energy tokens are generated and traded in a lower or higher amount than the actual energy impacting the physical grid by the overloading of the corresponding components, or even power outages, in extreme conditions [2]. Thus, mechanisms for creating and burning digital tokens need to be foreseen to keep them synchronized with the actual amount of energy generated.

Problems might occur due to the discretization of time in such a market, and due to the representation of energy consumption/production as digital tokens per time interval. Studies show that this might lead to excessive costs and not optimal schedules in the market-clearing results [3]. Alternative solutions are based on linear power trajectories, which represent the momentary electricity production that might reduce costs in a real-world system.

Moreover, the energy markets are usual organized as forward markets such as day-ahead markets. In this case, the trading session is divided into hours of the next day and the energy demand and production in the grid need to be matched for each respective hour. Participating entities place orders in advance, having then to try to live up to their promises [4], [5]. Thus, mechanisms for locking the energy tokens up to the moment of actual delivery of energy needs to be provided to make the settlement process feasible from both energy and financial perspective.

It is a desirable feature to enable the energy tokens locking by an escrow thus preventing the owner from spending them and ensuring the recipient that the tokens are safely stored until the energy transactions are settled based on actual monitored data. As it resulted from the state-of-the-art analysis done in section 2 nowadays solutions and standards for creating and managing tokens are not offering this feature. Thus, in this paper we define a Lockable ERC20 token that provides the option for an owner to lock some of its tokens, specifying a time-lock (after which the tokens can be unlocked) and an escrow party account (a third party, which can be a smart contract) responsible to unlock the tokens when certain conditions are met. To show its effectiveness a peer-to-peer energy trading scenario has been considered, in which the Lockable ERC20 token was used to digitize the surplus of energy of prosumers, the tokens being successfully locked up to the moment of energy transactions settlement considering the monitored data.

The rest of the paper is structured as follows: Section II presents the existing state of the art in assets digitization and token creation, Section III shows how the new Lockable ERC20 token has been implemented considering existing standards in the area, Section IV presents a peer to peer energy trading scenario in which the energy token is locked up to the point of settlement, while Section V concludes the paper.

## II. RELATED WORK

The blockchain-based distributed ledger technology had introduced the concept of digital tokens in business and economy allowing the tracking back of digital assets and fair allocation of revenue to the actual producers and service users that bring added value [6]. The are several types of tokens that can be used to comply with the regulations of different domains and for digitizing various physical assets [7]. The digital tokens are classified into three categories based on their purpose [8].

The payment tokens are cryptocurrencies that have no other functions, and they are used exclusively for payment fulfilment. Security or asset tokens correspond with equities, bonds, or derivatives and their purpose is to represent real-life assets. Also, there are utility tokens that provide access to the application and can be used to give users voting rights, rewards, or governance for their stakes. Though, in practice, the tokens are hybrids and have purposes from more than one category. The link between the different tokens' functions, the traded market, and their price on the market is studied in [9]. The tokens are categorized based on their functions and features. A fourth category, the *yield tokens* is identified, but these tokens often have other associated functionality and can be used as rewards for example in the proof of stake mechanism.

Tokens can also be divided into fungible and non-fungible [10], based on their ability to hold data. The non-fungible tokens are used for blockchain-traded rights to a digital asset (including for example the energy produced) and have become very popular lately.

Payment or exchange tokens are associated with electronic money [11] and their purpose is strictly related to payment and exchange, without other functionalities. They can be used to buy or sell goods and services which are not integrated on the platform. These kinds of tokens are usually associated with fungible tokens which are identical to one another, thus they are also interchangeable and can be divided [12]. Bitcoin and Ethereum are an example of fungible tokens. Bitcoin is one of the most popular cryptocurrencies which can be classified as a payment token, although it is also used as rewards for miners and reputation for users [13]. As was mentioned before, in practice tokens are most often hybrids. Bitcoin can be used as a payment tokens to buy services or goods which are external to the platform. Nowadays there are many fungibles digital token cryptocurrencies characterized by high security due to blockchain technology, immediate settlement, and fast transactions.

The asset tokens can have a variety of use cases in the digitalization of real-world assets with a design to make transactions more efficient [14-16]. They are non-fungible, unique, and can be used as proof of ownership over the asset. They are useful in addressing various problems of nowadays transaction systems such as centralization reducing the implication of third-party entities which manages the asset [17,18]. Currently, the transaction can be very complex with a lack of transparency and accountability. There are many steps involved, such as public listing, transactions on a secondary market, and asset management. Also, trading using alternative assets is even more complicated, involving many intermediaries. Using blockchain tokens to digitalize infrastructure or financial instruments has a positive impact on market transaction tracking, transparency, and auditability [19]. There is a legal contract without the involvement of a third party and the transaction is made almost instantly. After the transaction is included in a block and has several additional blocks as confirmations, it can be considered final. Thus, trading is made more resilient, tamper-proof and its accessibility is significantly improved.

Despite the advantages of having a blockchain market with digital tokens, there are some challenges related to the lack of financial regulations [20]. These uncertainties cause a lack of trust since security and stability are not officially covered from a legal point of view. Also, another issue to consider is how they can be adapted to be consistent with the off-chain assets. Even if the imposed requirements can lead to an increase in transaction costs, once these regulations are made, the enhanced liquidity of the asset remains an advantage [21], [22]. Nevertheless, blockchain technology can offer support for the tokenization of assets. It can be used to reduce trading and transactions time especially in the post-trade cycle [23]. The software artifacts need to deal with requirements related to security, heterogeneity, and usability, which are met using permission management, separating data from functionality, and access management [24]. The roles of different stakeholders who interact with the assets should be carefully defined and security schemes need to be imposed.

The impact that using blockchain technology can have on finances and different markets is increasingly studied [25-27]. In the case of shared ownership for an asset, multiple owners can use the same it based on a smart key they received. The challenge lies in representing the physical asset on-chain and shares the asset between all the owners. The second challenge is the peer-to-peer trading of the asset where blockchain technology is used to improve the market level traceability. Blockchain tokens can be used for energy trading [28], where a demurrage mechanism is used to deal with the decrease in the value of the tokens over time. The proposed mechanism is tested on a renewable energy market with simulated dynamics. Non-fungible tokens and blockchain can change the marketplace dynamics and affect the value chains by analysis of cryptocurrencies on the creative industry [29] (e.g., the digitization of art pieces).

To enable token creation and trading token standards have been proposed. They can offer additional functionalities for authentication and different utilities or simply for keeping a record of the holdings and transfer management. In the case of fungible tokens, the most relevant ones are ERC20 defined for Ethereum blockchain network [30], FA1.2 (TZIP-7) for Tezos [31], and Flow Fungible Token interface for Flow [32]. ERC20 is a contract standard interface that enables token creation, transfer operations, and approval for tokens to be spent by another account. The most used non-fungible Ethereum standard is ERC721 [33], Flow-NFT Standard for Flow [34], etc. ERC165, a Standard Interface Identifier that can be used to determine what interfaces a smart contract implements [35]. Finally, there are many token standards defined especially for the Ethereum blockchain network [36, 37] which can be developed to meet the application requirements.

### III. LOCKABLE ERC20 TOKEN CONTRACT

We have defined a Lockable ERC20 token contract to provide for an owner the option to lock some of its tokens, specifying a time-lock (after which the tokens can be unlocked) and an escrow party account (a third party, which can be a smart contract) responsible to unlock the tokens when certain conditions are met. It is built as an extension of the ERC20 standard defined by Ethereum for virtual asset representation that offers API implementation for tokens within Smart Contracts. Usually, the ERC20 standard is adopted to represent the tokens that the market participants offer as payment.

In addition to the functionalities available in ERC20, a safety mechanism was necessary to ensure that the participants own the tokens they offer, and they will not be able to spend them for a period, thus the ERC20 implements a time-lock governed escrow between the owner (spending account) and the recipient (receiving account). This meets the requirements of various business scenarios (e.g., energy trading domain), where it is desired that the tokens be locked by an escrow, thus preventing the owner from spending them, and ensuring the recipient that the tokens are safely locked. When meeting certain given business conditions, for example after transactions settlement the tokens are transferred to the recipient or back to the owner their wallets being settled.

Each ERC20 contract is identified by a name and a symbol which can be set only once during construction and provides functionalities for token management such as token transfer and approval for tokens to be used by a third party. ERC20 contract uses the SafeMath library for token operations. It has internal mapping **_balances** that map each account address to its token balance. For allowance (permission given to an address to spend tokens owned by another address) ERC20 contract uses a double mapping **_allowance** that links the owner with the spender and stores the amount approved to be spent. Figure 1 shows the methods implemented by the ERC20 Contract to provide these functionalities methods that will be adapted and extended to bring the Lockable feature desired in our case.

**ERC20 Functions**
1: **totalSupply**() public view returns (uint256)
2: **balanceOf**(address _owner) public view returns (uint256 balance)
3: **transfer**(address _to, uint256 _value) public returns (bool success)
4: **transferFrom**(address _from, address _to, uint256 _value) public returns (bool success)
5: **approve**(address _spender, uint256 _value) public returns (bool success)
6: **increaseAllowance**(address spender, uint256 addedValue)
7: **decreaseAllowance**(address spender, uint256 subtractedValue) public virtual returns (bool)
8: **allowance**(address _owner, address _spender) public view returns (uint256 remaining)

Fig. 1. ERC20 functions relevant to implement the new Lockable feature

In our solution, a lock is added over a given quantity of tokens, by specifying: the locker (the transaction signing address), the number of tokens, the locking period, and the unlocking responsible (third party escrow).

The *Token Locker* is a market participant which designates a smart contract to act on behalf of him, thus registering orders and locking tokens according to some predefined set of rules. Such a contract needs the rights to lock tokens from its owner account being possible through the allowance functionality, which permits the owner to give locking and transferring rights for a given quantity of tokens to the designated smart contract.

The *Locking Period* is the time interval during which the tokens are locked (i.e., usually the duration of the market session). When the orders are matched and the trades are established, the tokens are released from their lock, thus the escrow party can check the order status and proceed with the unlocking phase.

The *Unlocking Responsible* (Third-party Escrow) is usually the market operator. After the matching phase, the market contract holds all the necessary information to be able to unlock the tokens and transfer them. In case an order has not been matched, the tokens return to the initial owner, otherwise, the escrow will initiate a transfer to the recipient (the selling party that has provided the asset).

To provide these additional functionalities, LockableERC20 (LERC20) extends the ERC20 contract implementation. The evidence of a locked balance for each account is kept using two additional mappings. The *_locked* mapping (see Figure 2 - line 3) links the address of the one responsible for that lock with the locker address (the locker can be the owner or an address that received approval from the owner to lock) and stores a *LockedEntity* data structure (Figure 3). The *_lockedBalance* mapping stores the locked balance for each account (Figure 2 - line 4).

**Smart Contract:** LockableERC20
1: **State:**
2:     **address** tokens_owner
3:     MAP (**address** unlocking_address, MAP(**address** locking_address, **LockedEntity** lockedEntity)) **_locked**
4:     MAP (**address** account_address, **uint256** balance) **_lockedBalance**
5: *Constructor*:
6:     **Input:** msg.sender, amount
7:     **Output:** -
8:     **Begin:**
9:         ERC20("LockableERC20", "LERC20")
10:        tokens_owner ← msg.sender
11:        mint(tokens_owner, amount)

Fig 2. LERC20 state and constructor

**Library:** LockedEntity
1: **Struct:** _Locked Entity
2:     **Members:**
3:         **address** owner
4:         **uint256** amount
5:         **uint** blockNo
6:         **bool** isActive

Fig 3. LockedEntity data structure

As was mentioned before, the locker can be different from the one that owns the tokens. To be consistent with the transfer operation, two methods were added: *lock* and *lockFrom*. Both methods use _lock method (see Figure 4) that updates (lines 12-15) or creates (line 17) the lock entity between the one responsible for unlocking (*unlockingAddress*) and the locker. The block number for the end of the session is computed by adding the *noBlocks* (session duration in a number of blocks) to the current block number (lines 15, 17). Then, it gives rights to the unlocking address to transfer tokens in the name of the owner (line 20) and increases the locked balance of the owner (line 21).

**Smart Contract:** LockableERC20
1: **Function** _lock
2: **Input:** block.number, owner, locker, amount, noBlocks, unlockingAddress
3: **Output:** -
4: **Modifiers:** private
5: **Begin:**
6:     **If** _locked[unlockingAddress][locker].isActive
7:         _locked[unlockingAddress][locker].amount.add(amount)
8:         _locked[unlockingAddress][locker].blockNo ←
9:             max(_locked[unlockingAddress][locker].blockNo, block.number+noBlocks)
10:    **Else**
11:        _locked[unlockingAddress][locker] ← new LockedEntity
                    (owner, amount, block.number + noBlocks, *True*)
12:    **End If**
13:    **Emit** LockEvent
14:    _approve(owner, unlockingAddress, allowance(owner, unlockingAddress) + amount)
15:    _lockedBalance[owner] ← lockedBalance[owner] + amount
16: **End**

17: **Function** *lock*

Fig 4. LERC20 _lock function

Also, the _lock function emits a *Lock event* used to keep track of all lock related operations (see Figure 5). Status is 0 for lock, 1 for unlock and 2 for unlock without transfer.

| Lock Event |
|---|
| 1: **Event:** Lock |
| 2:     **address indexed** owner |
| 3:     **address** lockingAddress |
| 4:     **uint256** amount |
| 5:     **uint** noBlocks |
| 6:     **address indexed** unlockingAddress |
| 7:     **uint8** status |
| 8:     **uint indexed** timestamp |
| 9:     **uint** hour |
| 10:     **uint** min |

Fig 5. LERC20 Lock Event

The locking operation (see Figure 6) requires that the account has enough available unlocked balance (line 7). If this condition is met, _lock internal function is called (line 8) and in this case, the owner field is the same as the locker account. It was also needed to allow someone else (in this case a contract owned by the participant) to lock tokens on behalf of the owner. The unlocked balance for the owner account is checked (line 16) and then the allowance is decreased using _approve internal function of the ERC20 contract. If the locked amount exceeds the allowance, this operation will fail. If all requirements are met, the internal _lock function is called (line 19). In this case, the locker is the *msgSender()* but the owner's account address is given as a parameter.

| **Smart Contract:** LockableERC20 |
|---|
| 1: **Function** *lock* |
| 2:   **Input:** msg.sender, amount, noBlocks, unlockingAddress |
| 3:   **Output:** - |
| 4:   **Modifiers:** public |
| 5:   **Begin:** |
| 6:     account ← msg.sender |
| 7:     **Requires** unlockedBalanceOf(account) >= amount |
| 8:     _lock(account, account, amount, noBlocks, unlockingAddress) |
| 9:   **End** |
| 10: **Function** *lockFrom* |
| 11:   **Input:** msg.sender, owner, amount, noBlocks, unlockingAddress |
| 12:   **Output:** - |
| 13:   **Modifiers:** public |
| 14:   **Begin:** |
| 15:     locker ← msg.sender |
| 16:     **Requires** unlockedBalanceOf(owner) >= amount |
| 17:     **Requires** allowance(owner, locker) >= amount |
| 18:     _approve(owner, locker, allowance(owner, locker) - amount) |
| 19:     _lock(owner, locker, amount, noBlocks, unlockingAddress) |
| 20:   **End** |

Fig 6. LERC20 – lock and lockFrom operations

The unlock operation (see Figure 7) unlocks a partial amount (line 8) or deactivates the *LockEntity* entry from _locked mapping (line 11), decreases the *lockedBalance* for the owner account (line 13), and then, initiates a *transferFrom* operation from the owner account to recipient with the specified amount (line 14). For this operation, *verifyLock* (line 7) method is used to make sure that the lock exists, and it can be unlocked (the session ended).

| **Smart Contract:** LockableERC20 |
|---|
| 1: **Function** *unlockTransfer* |
| 2:   **Input:** msg.sender, owner, amount, recipient, lockingAddress |
| 3:   **Output:** - |
| 4:   **Modifiers:** public |
| 5:   **Begin:** |
| 6:     unlockingAddress ← msg.sender |
| 7:     **Requires** verifyLock(owner, amount, unlockingAddress, block.number) is *True* |
| 8:     **If** amount < _locked[unlockingAddress][lockingAddress].amount |
| 9:       _locked[unlockingAddress][lockingAddress].amount ← _locked[unlockingAddress][lockingAddress].amount – amount |
| 10:     **Else** |
| 11:       _locked[unlockingAddress][lockingAddress].isActive = false |
| 12:     **End If** |
| 13:     _lockedBalance[owner] ← _lockedBalance[owner] – amount |
| 14:     transferFrom(owner, recipient, amount) |
| 15:     **Emit** UnlockTransferEvent |
| 16:   **End** |

Fig 7. LERC20 unlockTransfer function

Also, we have defined an *unlockWithoutTransfer* method similar to one but without the transfer operation, thus the tokens remaining in the account of the initial owner.

Overriding of ERC20 *transfer* and *transferFrom* functions was necessary to ensure that transfer is allowed only for unlocked balance. Also, the allowance can be increased only for the tokens that are not locked (Figure 8).

| **Smart Contract:** LockableERC20 |
|---|
| 1: **Function** *transfer* |
| 2:   **Input:** msg.sender, recipient, amount |
| 3:   **Output:** - |
| 4:   **Modifiers:** public override |
| 5:   **Begin:** |
| 6:     **Requires** amount <= unlockedBalanceOf(msg.sender) |
| 7:     _transfer(msg.sender, recipient, amount) |
| 8:   **End** |
| 9: **Function** *transferFrom* |
| 10:   **Input:** msg.sender, sender, recipient, amount |
| 11:   **Output:** - |
| 12:   **Modifiers:** public override |
| 13:   **Begin:** |
| 14:     **Requires** amount <= unlockedBalanceOf(sender) |
| 15:     **Requires** allowance(sender, msg.sender) >= amount |
| 16:     _transfer(msg.sender, recipient, amount) |
| 17:     _approve(sender, msg.sender, allowance(sender, msg.sender) - amount) |
| 18:   **End** |
| 19: **Function** *increaseAllowance* |
| 20:   **Input:** msg.sender, spender, amount |
| 21:   **Output:** - |
| 22:   **Modifiers:** public override |
| 23:   **Begin:** |
| 24:     **Requires** amount <= unlockedBalanceOf(msg.sender) |
| 25:     _approve(msg.sender, spender, allowance(msg.sender, spender) + amount) |
| 26:   **End** |
| 27: **Function** *decereaseAllowance* |
| 28:   **Input:** msg.sender, spender, amount |
| 29:   **Output:** - |
| 30:   **Modifiers:** public override |
| 31:   **Begin:** |
| 32:     **Requires** amount <= allowance(msg.sender, spender) |
| 33:     _approve(msg.sender, spender, allowance(msg.sender, spender) - amount) |

34:    **End**

Fig 8. LERC20 overriding transfer and allowance methods

## IV. VALIDATION RESULTS

The validation for the Lockable ERC20 contract was made in the context of the peer-to-peer energy market the settlement being done at the end of the market session. The LERC20 token is used to represent the surplus of energy of a peer prosumer submitted as an energy offer on the market and provides an insurance mechanism for the tokens involved in market energy transactions (i.e., a pairing between an energy sell offer and an energy buy bid).

When a sell energy offer is registered on the market, a number of tokens equal with the

$$No_{Tokens} = E_{price} * E_{amount} \quad (1)$$

are locked from the owner's account. In (1) $E_{price}$ and $E_{amount}$, represent the amount of energy and price associated with the sell energy offer. In this was tokens are locked in each energy offer successfully submitted during a market session until the energy and financial settlement is being conducted. During delivery, if the produced energy is lower than the one promised in the sell offer only the corresponding number of tokens is transferred from the seller account to the buyer account. Only the tokens for the delivered energy are unlocked and remains returned to the owner's account. In the case of buy energy bids, the tokens which are involved are also locked guarantying the solvability of the energy buyer. When the energy is successfully received, the corresponding tokens are unlocked and transferred from the buyer to the seller account.

Thus, it ensures that the energy buyer is not able to spend the tokens and, at the same time, that he has the respective amount to pay for the delivery. Tokens are safely kept in the owner account until settlement and the transfer is made only, if necessary, at the end of the session. The time for which the tokens are locked is specified in chain blocks when the lock is created. Also, a contract that handles the commitments is assigned to be the third-party escrow responsible for the unlock operation.

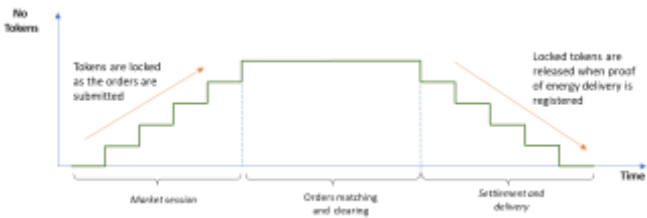

Fig 9. P2P market phases with lockable tokens

In Figure 9 we present the phases of the P2P energy market when the locking mechanism is put in place. Once the market session opens, and the market is ready to receive orders, the locking phase begins. Each placed order will also be reflected in a lock being acquired. The number of locks will thus increase, until the end of the session, when the algorithm will run against the placed bid/sell orders. Once the trades are computed, and the delivery process starts between the matched pairs, the locks are also released with each trade that is being settled. Thus, the number of locks will start to decrease until the end of the delivery period when they should all be released.

The number of possible transactions on the blockchain is given by the gas limit for a block and the gas consumed by the transaction. Thus, any additional operation to a transaction like in this case the token locking may affect the scalability. It will increase the gas cost and implicitly will lead to a decrease in the number of transactions that can be performed in a period.

To evaluate the impact on the scalability brought by our tokens locking solution (i.e., the additional lock and unlock operations), two scenarios were studied on the peer-to-peer market scenario (see Figure 10). We have considered the current gas limit of 15 million for a block on the Ethereum blockchain network and that a block is mined around every 15 seconds. The number of possible buy and sell energy order transactions were computed both with and without the lock operation imposed by our solution.

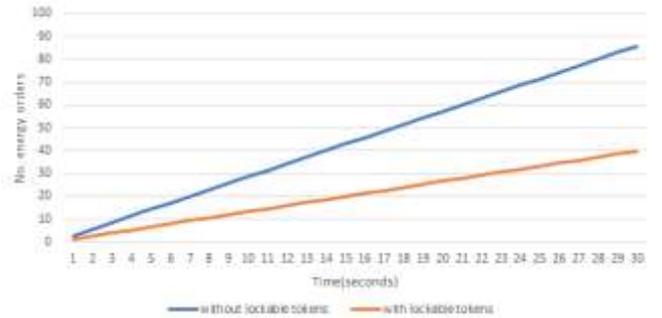

Fig 10. Number of orders over time

First, the transaction for energy bids and offers placement in the market session was made without the lock operation. According to the gas cost of the transaction (gas used: 348774), 42 orders without the lock operation can be placed on the market every 15 seconds. When the lock operation was added to the place order transaction the gas used for the transaction increases (gas used: 748565), thus around 20 orders on the market are allowed every 15 seconds.

Considering the potential length of a market session of about 4 hours the impact of the lock operation gas consumption is relatively low, but at the same time, it provides assurances regarding the tokens involved in market transactions. After the order is placed, the owner is not able to spend the tokens until the promised energy is delivered and the responsible third-party escrow unlocks them at the end of the market session. Thus, the situations in which the tokens necessary for a financial settlement are spent are avoided.

## V. CONCLUSION

In this paper, we have presented a Lockable ERC20 token that allows the locking tokens for a specific time or until certain business conditions are met. This feature is desirable in decentralized energy management scenarios in which the blockchain transactions represent promises of future energy delivery. To validate the proposed approach, we have considered the peer-to-peer energy trading scenario the Lockable ERC20 token being effective in used to digitize the surplus of energy of prosumers and locking the energy tokens committed in transactions up to the actual delivery.

For further work, we plan to use the lockable token to implement trustable and secure mechanisms for energy assets monetization, including the energy flexibility from heterogeneous sources beyond electricity such as thermal, comfort services, etc.

ACKNOWLEDGMENT

This work has been conducted within the BRIGHT project grant number 957816 funded by the European Commission as part of the H2020 Framework Programme and it was partially supported by a grant of the Romanian Ministry of Education and Research, CNCS/CCCDI–UEFISCDI, project number PN-III-P3-3.6-H2020-2020-0031.